\documentclass[
    ,final            % use final for the camera ready runs
  ]
  {aipproc}

\layoutstyle{6x9}

\begin{document}
 
\title{Superposition rules and second-order differential equations}


\author{J.F. Cari\~nena}{
  address={Departamento de F\'isica Te\'orica, Universidad de Zaragoza, Pedro Cerbuna 12, 50.009, Zaragoza, Spain}
}

\author{J. de Lucas}{
  address={Institute of Mathematics, Polish Academy of Sciences, ul. \'Sniadeckich 8, P.O. Box 21, 00-956, Warszawa, Poland}
}

\begin{abstract} The main purpose of this work is to introduce and analyse some generalizations of diverse superposition rules for first-order differential equations to the setting of second-order differential equations. As a result, we find a way to apply the theories of Lie and quasi-Lie systems to analyse second-order differential equations. In order to illustrate our results, several second-order differential equations appearing in the physics and mathematical literature are analysed and some superposition rules for these equations are derived by means of our methods.
\end{abstract}

\maketitle

%%%%%%%%%%%%%%%%%%%%%%%%%%%%%%%%%%%%%%%%%%%%
%% MAINMATTER
%%%%%%%%%%%%%%%%%%%%%%%%%%%%%%%%%%%%%%%%%%%%
\section{Motivation}
It is a known fact that every linear homogeneous system of first-order differential equations admits a {\it linear superposition rule}, namely, its general solution can be written in terms of a linear combination of a family of linearly independent particular solutions and a set of constants to be related to initial conditions. Nevertheless, it is not so well-known that these systems can be viewed as a particular example of a larger class of first-order systems, the so-called {\it Lie systems}, that admit a `more general' type of superposition rule which, for instance, need not be a linear combination of particular solutions \cite{CGM00,CGM07}. Additionally, a more general type of superposition rule has been found for the denominated {\it quasi-Lie systems} \cite{CGL09}, which include, as particular cases, Lie systems. 

Taking now into account that linear homogeneous systems of second-order differential equations also admit a certain type of linear superposition rule, it is natural to ask ourselves what kind of systems of second-order differential equations admit their general solution to be obtained, in a more general way, in terms of certain families of particular solutions and sets of constants. As witnessed by some previous works \cite{CGL09,RSW97,SecOrd,CLR08}, the analysis of this question may lead to finding new insights and results in the study of many problems of mathematical and physical interest. 

Motivated by the above facts, the main aim of this work is to present some results concerning the study of superposition rules for second-order differential equations. More specifically, we next provide several definitions of these new types of superposition rules along with some results describing special kinds of second-order differential equations admitting them. As a direct consequence of our results, it turns out that the theories of Lie and quasi-Lie systems can be applied to analyse second-order differential equations. Finally, we carry out various applications of our results to the analysis of a number of second-order differential equations appearing in the physics and mathematical literature, paying special attention to second-order Riccati equations.

\section{Fundamentals}
Let us here briefly recall some fundamental notions to be used throughout this work. For a detailed description of these and other related topics, see \cite{CGM00,CGL09,SecOrd}.

A {\it superposition rule} for a system of differential equations 
\begin{equation}\label{sys}
\frac{dx^i}{dt}=X^i\left(t,x\right),\qquad i=1,\ldots,n,\\
\end{equation}
defined on ${R}^n$ is a map $\Phi:({R}^{n})^m\times{R}^n\rightarrow {R}^n$ such that the general solution, $x(t)$, of the system can be written, at least locally, as
$$
x(t)=\Phi(x_{(1)}(t),\ldots,x_{(m)}(t),k_1,\ldots,k_n),
$$
in terms of any `generic' set of particular solutions $x_{(1)}(t),\ldots,x_{(m)}(t)$ and a set of constants $k_1,\ldots,k_n$ to be related to the initial conditions of the system. 

The characterization of those systems of the form (\ref{sys}) admitting a superposition rule is due to the Norwegian mathematician Sophus Lie. In modern geometric terms, Lie's characterization (the today called {\it Lie Theorem}) states that a system of the form (\ref{sys}) admits a superposition rule, i.e. it is a {\it Lie system}, if and only if its associated $t$-dependent vector field, i.e. $X(t,x)=\sum_{i=1}^nX^i(t,x)\partial/\partial x^i$, can be cast into the form
\begin{equation}\label{DecLieSys}
X(t,x)=\sum_{\alpha=1}^rb_\alpha(t)X_\alpha(x),
\end{equation}
where $X_1,\ldots,X_r$ are a set of vector fields on ${R}^n$ spanning a finite-dimensional Lie algebra of vector fields, the so-called {\it Vessiot-Guldberg Lie algebra}.

In a more general picture, the $t$-dependent vector field associated with a system of the form (\ref{sys}) might be cast into the form
\begin{equation}\label{DecQuaLieSys}
X(t,x)=\sum_{\alpha=1}^sc_\alpha(t)Y_\alpha(x),
\end{equation}
where the vector fields $Y_\alpha$ do not need to close a Lie algebra, but they must span a finite-dimensional vector space $V_2$ admitting a subspace $W$ such that $[W,W]\subset W$ and $[W,V_2]\subset V_2$. In such a case, the pair $W,V_2$ is said to form a {\it quasi-Lie scheme $S(W,V_2)$} and there exists a group of $t$-dependent transformations, the so-called {\it group of the scheme}, $\mathcal{G}(W)$, whose elements transform our initial system (\ref{sys}) into new ones related to $t$-dependent vector fields of the form $X'(t,x)=\sum_{\alpha=1}^sc_\alpha'(t)Y_\alpha(x)$, see \cite{CGL09}. Moreover, if for a certain $t$-dependent transformation of $\mathcal{G}(W)$, the $t$-dependent vector field $X'(t,x)$ can be cast into a form similar to (\ref{DecLieSys}), the initial system is said to be a {\it quasi-Lie system with respect to the scheme $S(W,V_2)$}. In this case, it can be proved that the general solution of the initial system can be written as $x(t)=\Phi_t(t,x_{(1)}(t),\ldots,x_{(m)}(t),k_1,\ldots,k_n)$ in terms of any `generic' set of particular solutions $x_{(1)}(t),\ldots,x_{(m)}(t)$ and a set of constants $k_1,\ldots,k_n$. In other words, system (\ref{sys}) is said to admit a {\it $t$-dependent superposition rule} $\Phi_t:{R}\times({R}^{n})^{m}\times{R}^n\rightarrow{R}^n$.

\section{Superposition rules and SODEs}
Motivated by the recent works studying second-order differential equations from the point of view of the theory of Lie systems \cite{SecOrd,CLR08,CLR07}, it turns out that the appropriate definition of superposition rule for second-order differential equations must be as follows.

Given a system of second-order differential equations
\begin{equation}\label{SecOr}
\frac{d^2x^i}{dt^2}=F^i\left(t,x,\frac{dx}{dt}\right),\qquad \qquad\qquad i=1,\ldots,n,
\end{equation}
on ${R}^n$, we say that it admits a {\it superposition rule} if there exists a mapping $\bar \Phi:({T}{R}^{n})^m\times{R}^{2n}\rightarrow {R}^n$ such that the general solution, $x(t)$, of the system can be written, at least locally, as
$$
x(t)=\bar \Phi\left(x_{(1)}(t),\frac{dx_{(1)}}{dt}(t)\ldots,x_{(m)}(t),\frac{dx_{(m)}}{dt}(t),k_1,\ldots,k_{2n}\right),
$$
in terms of any `generic' set of particular solutions $x_{(1)}(t),\ldots,x_{(m)}(t)$, their derivatives with respect to the independent variable $t$, and a set of constants $k_1,\ldots,k_{2n}$ to be related to the initial conditions. 
	      
A useful concept in order to recognize second-order differential equations admitting a superposition rule is the {SODE Lie system} notion. Let us define this concept. Given a second-order differential equation (\ref{SecOr}), we say that it is a {\it SODE Lie system} if its associated first-order system
\begin{equation}\label{AsoFir}
\left\{\begin{array}{l}
\frac{dx^i}{dt}=v^i,\\
\frac{dv^i}{dt}=F^i\left(t,x,v\right),\\
\end{array}\right.\qquad\qquad\qquad i=1,\ldots,n,
\end{equation}
is a Lie system.
    
The interest on the SODE Lie system concept is motivated by the following result, whose demonstration follows straightforwardly from \cite[Proposition 1]{SecOrd}.

{\bf Proposition 1.} {\it Every SODE Lie system of the form (\ref{SecOr}) admits a superposition rule $\bar \Phi=\pi\circ\Phi$, where $\Phi:({T}{R}^n)^{m}\times {R}^{2n}\rightarrow {T}{R}^n$ is a superposition rule for (\ref{AsoFir}) and $\pi:{T}{R}^n\rightarrow{R}^n$ is the projection map related to the tangent bundle ${T}{R}^n$.}	       
      
Recently, the theory of quasi-Lie schemes \cite{CGL09} introduced a new type of superposition rule for systems of first-order differential equations that we next define. A system (\ref{SecOr}) is said to admit a {\it $t$-dependent superposition rule} if there exists a map $\bar \Phi_t:{R}\times({T}{R}^n)^m\times{R}^{2n}\rightarrow{R}^n$ such that its general solution can be cast into the form
$$
x(t)=\bar\Phi_t\left(t,x_{(1)}(t),\frac{dx_{(1)}}{dt}(t),\ldots,x_{(m)}(t),\frac{dx_{(m)}}{dt}(t),k_1,\ldots,k_{2n}\right),
$$
in terms of any `generic' set of particular solutions $x_{(1)}(t),\ldots,x_{(m)}(t)$, their derivatives with respect to the independent variable $t$, and a set of constants $k_1,\ldots,k_{2n}$ to be related to the initial conditions. 
	
In a similar way as Proposition 1 shows the existence of superposition rules for SODE Lie systems, it can be proved the following result that ensures the existence of $t$-dependent superposition rules for a more general class of systems of second-order differential equations.
%\vskip 0.5cm

{\bf Proposition 2.} {\it Every second-order system (\ref{SecOr}), whose associated first-order system (\ref{AsoFir}) is a quasi-Lie system with respect to some quasi-Lie scheme, admits a $t$-dependent superposition rule of the form $\bar\Phi_t=\pi\circ\Phi_t$, where $\Phi_t$ is a $t$-dependent superposition rule associated with the quasi-Lie system (\ref{AsoFir}).}

\section{Applications}
Let us briefly explain the application of the previous theoretical results to the analysis of some second-order differential equations appearing in the mathematical and physics literature \cite{CRS05}. Our first aim is concerned with analysing the second-order differential equation
\begin{equation}\label{PI}
\frac{d^2x}{dt^2}+3x\frac{dx}{dt}+x^3=f(t),
\end{equation}
related to the study of B\"acklund transformations for the Sawada-Kotera PDE and appearing in the study of the so-called Riccati chains \cite{CRS05,GL99}. More specifically, we pretend to prove that equation (\ref{PI}) is a SODE Lie system and to describe one of its associated superposition rules by means of Proposition 1. In order to do so, note that equation (\ref{PI}) is such that its associated first-order system (\ref{AsoFir})
describes the integral curves of the $t$-dependent vector field $X_t=X_1+f(t)X_2$, where $X_1=v\partial_x-(3xv+x^3)\partial_v$ and $X_2=\partial_v$. It can be proved that the vector fields $X_1$ and $X_2$ span, along with the vector fields
$$
\begin{array}{lll}
&X_3=-\partial_x+3x\partial_v,\,\, &X_4=x\partial_x-2x^2\partial_v,\\
&X_5=(v+2x^2)\partial_x-x(v+3x^2)\partial_v,\,\, &X_6=2x(v+x^2)\partial_x+2(v^2-x^4)\partial_v,\\
&X_7=\partial_x-x{\partial_v},\,\,&X_8=2x\partial_x+4v\partial_v,
\end{array}
$$
an eight-dimensional Lie algebra of vector fields $V$ isomorphic to ${sl}(3,{R})$, see \cite{SecOrd}. It follows that equation (\ref{PI}) is a SODE Lie system and, in virtue of Proposition 1, it admits a superposition rule. 
Following a method to obtain superposition rules described in \cite{CGM07}, it can be proved that the general solution for system (\ref{PI}) can be written as
\begin{equation}\label{SupFirstRicc}
x(t)=\frac{x_2(t)F_{431}-G_{3124}k_2-G_{2134}k_1+x_3(t)F_{421}k_1k_2}{F_{431}+(F_{124}-F_{324})k_1+(F_{412}-F_{312})k_2+F_{421}k_1k_2},
\end{equation}
where $G_{abcd}$ and $F_{abc}$, with $a,b,c,d=1,\ldots,4$, are certain functions depending only on the particular solutions $x_1(t),\ldots,x_4(t)$ and their derivatives, and $k_1,k_2$ are real constants. For a detailed description of this result, we refer the reader to \cite{SecOrd}.

Apart from system (\ref{PI}), many other second-order differential equations are SODE Lie systems that, when transformed into first-order ones, are related to Lie systems whose associated $t$-dependent vector fields are described by linear combinations of the form (\ref{DecLieSys}) of vector fields in $V$. Among these SODE Lie systems, we can single out the equations
$$
\frac{d^2x}{dt^2}+3x\frac{dx}{dt}+x^3+g(t)\left(\frac{dx}{dt}+x^2\right)+h(t)x+j(t)=0,
$$
appearing, for instance, in the study of differential equations with maximal number of Lie symmetries. As a consequence of Proposition 1, all the members of this family admit the same superposition rule as equation (\ref{PI}) and therefore their general solutions can also be cast into the form (\ref{SupFirstRicc}).

Let us sketch now how Proposition 2 can be used to derive a $t$-dependent superposition rule for the second-order Riccati equations \cite{CRS05,GL99} of the form
\begin{equation}\label{NLe}
\frac{d^2x}{dt^2}+(b_0(t)+b_1(t)x)\frac{dx}{dt}+a_0(t)+a_1(t)x+a_2(t)x^2+a_3(t)x^3=0,
\end{equation}
with $a_3(t)>0$, $a_3(0)=1$, $b_1(t)=3\sqrt{a_3(t)}$, and $b_0(t)=a_2(t)/\sqrt{a_3(t)}-d{a}_3/dt(t)/(2a_3(t))$. For a full description of the following techniques, see \cite{SecOrd}. 

In order to apply Proposition 2 to equation (\ref{NLe}), it is necessary to prove that the system
\begin{equation}\label{FOSOR}
\left\{
\begin{array}{ll}
&\frac{dx}{dt}=v,\\
&\frac{dv}{dt}=-(b_0(t)+b_1(t)x)v-a_0(t)-a_1(t)x-a_2(t)x^2-a_3(t)x^3,
\end{array}\right.
\end{equation}
obtained by adding the variable $v=dx/dt$ to the equation (\ref{NLe}), is a quasi-Lie system. Consider the following set of vector fields
$$Y_1=v{\partial_x},\,\, Y_2=v{\partial_v},\,\, Y_3=xv{\partial_v},\,\, Y_4={\partial_v},\,\, Y_5=x{\partial_v},\,\, Y_6=x^2{\partial_v},\,\, Y_7=x^3{\partial_v},\,\, Y_8=x{\partial_x},$$spanning a linear space of vector fields $V_2=\langle Y_1,\ldots,Y_8\rangle$ and define $W=\langle Y_2,Y_8\rangle$. The linear space $W$ is a two-dimensional Abelian Lie algebra of
vector fields and it can be proved that $[W,V_2]\subset V_2$. Hence, the pair $W,V_2$ forms a quasi-Lie scheme $S(W,V_2)$. Note that system (\ref{FOSOR}) describes the integral curves of the $t$-dependent vector field 
$$
Y_t=Y_1-b_0(t)Y_2-b_1(t)Y_3-a_0(t)Y_4-a_1(t)Y_5-a_2(t)Y_6-a_3(t)Y_7,
$$
of the form (\ref{DecQuaLieSys}) in terms of the vector fields of $V_2$. Consequently, the so-called group of the scheme $\mathcal{G}(W)$ associated with $S(W,V_2)$ can be used to transform the system determined by $Y_t$ into a new system determined by a $t$-dependent vector field taking values in $V_2$, see \cite[Proposition 1]{CGL09}. In particular, among the elements of $\mathcal{G}(W)$, it is easy to find that the $t$-dependent change of variables $\bar x=x,$ $\bar v=(a_3(t))^{-1/2}dx/dt$, converts system (\ref{FOSOR}) into the Lie system describing the integral curves of the $t$-dependent vector field
$$
X_t=\sqrt{a_3(t)}X_1-\frac{a_0(t)}{\sqrt{a_3(t)}}X_2-\frac{a_1(t)}{2\sqrt{a_3(t)}}(X_3+X_7)-\frac{a_2(t)}{4\sqrt{a_3(t)}}(X_8-2X_4).
$$
In consequence system (\ref{FOSOR}) is a quasi-Lie system with respect to the scheme $S(W,V_2)$ and  Proposition 2 ensures the existence of a $t$-dependent superposition rule for every member of the family (\ref{NLe}). More specifically, the superposition rule for system $X_t$ gives rise to a $t$-dependent superposition rule for the system associated with $Y_t$, by inverting the previous $t$-dependent change of variables \cite[Section 4]{CGL09}. From this superposition rule, it follows straightforwardly (cf. \cite{SecOrd}) that the general solution for every member of the family (\ref{NLe}) can be cast into the form 
\begin{equation}\label{SupRicc}
x(t)=\frac{x_2(t)\widetilde F_{431}-\widetilde G_{2134}k_1-\widetilde G_{3124}k_2+x_3(t)\widetilde F_{421}k_1k_2}{\widetilde F_{431}+(\widetilde F_{124}-\widetilde F_{324})k_1+(\widetilde F_{412}-\widetilde F_{312})k_2+\widetilde F_{421}k_1k_2},
\end{equation}
where $\widetilde{F}_{abc}$ and $\widetilde{G}_{abcd}$, with $a,b,c,d=1,\ldots,4$, are certain $t$-dependent functions depending on any generic family of particular solutions $x_1(t),\ldots,x_4(t)$ and their derivatives, and $k_1, k_2$ are two real constants.
\section{Conclusions and Outlook}
Several types of superposition rules for systems of second-order differential equations have been introduced, and some classes of these systems have been proved to admit these new superposition rules. As an application, we have derived $t$-dependent and $t$-independent superposition rules for various second-order differential equations appearing in the physics and mathematical literature. The definitions and methods showed here  seem to be generalizable to the setting of higher-order systems of differential equations, where they could provide, for instance, a method to study the differential equations of the so-called {\it Riccati hierarchy} appearing in the study of B\"acklund transformations of several PDEs \cite{GL99}. We aim to investigate these and other related topics in future works.  

%%%%%%%%%%%%%%%%%%%%%%%%%%%%%%%%%%%%%%%%%%%%%%%%
%% BACKMATTER
%%%%%%%%%%%%%%%%%%%%%%%%%%%%%%%%%%%%%%%%%%%%%%%%

\begin{theacknowledgments}
  Partial financial support by research projects MTM2009-11154, MTM2009-08166-E, and E24/1 (DGA)
are acknowledged. 
\end{theacknowledgments}



\bibliographystyle{aipproc}   % if natbib is available

\begin{thebibliography}{99}
\bibitem{CGM00}
J.F. Cari\~nena,  J. Grabowski and G. Marmo,
{\it Lie--Scheffers systems: a geometric approach,}
Bibliopolis, Naples, 2000.
\bibitem{CGM07}
J.F. Cari{\~n}ena, J. Grabowski, and G. Marmo, \emph{Rep. Math. Phys.} {\bf 60}, 237--258 (2007).
\bibitem{CGL09}
J.F. Cari{\~n}ena, J. Grabowski, and J. de Lucas, \emph{ J. Phys. A} {\bf 42}, 335206 (2009).
\bibitem{RSW97}
C. Rogers, W.K. Schief, and P. Winternitz,
\emph{J. Math. Anal. Appl.} {\bf 216}, 246--264 (1997).
\bibitem{SecOrd}
J.F. Cari\~nena and J. de Lucas, \emph{J. Geom. Mech.} {\bf 3}, 1--22 (2011).
\bibitem{CLR08}
J.F. Cari{\~n}ena, J. de Lucas, and M.F. Ra{\~n}ada, \emph{SIGMA Symmetry Integrability Geom. Methods Appl.} {\bf 4}, 031 (2008).
\bibitem{CLR07}
J.F. Cari{\~n}ena, J. de Lucas, and M.F. Ra\~nada, ``Nonlinear superpositions and Ermakov systems'',  in \emph{Differential Geometric Methods in Mechanics and Field Theory},  edited by F. Cantrijn, M. Crampin, and B. Langerock, Academia Press, Prague, 2007, pp. 15--33. 
\bibitem{CRS05}
J.F. Cari{\~n}ena, M.F. Ra{\~n}ada, and M. Santander, \emph{J. Math. Phys.} {\bf 46}, 062703 (2005).
\bibitem{GL99}
A.M. Grundland and D. Levi, \emph{J. Phys. A} {\bf 32}, 3931--3937 (1999).
\end{thebibliography}

\end{document}